\documentclass[10pt,a4paper]{article}
\usepackage{graphicx}
\usepackage{graphics}
\title{Elliptically Symmetric Polarized Beams}
\author{Omar El Gawhary\\
Dipartimento di Fisica and Istituto Nazionale per\\ la
Fisica della Materia,
Universit\`a ``Roma Tre''\\
Via della Vasca Navale 84, I-00146 Rome, Italy}
\date{}
\begin{document}
\maketitle
\date{}
\begin{abstract}
We study the free-propagation features of an optical field endowed with a non-uniform polarization
  pattern with elliptical symmetry. The fields derived in this way
are called Elliptically Symmetric Polarized Beams (ESPB for short). Some properties
 of those fields are analysed. Moreover, it is 
shown how it is possible to obtain such light beams by applying the result
to Bessel-Gauss beams.
\end{abstract}



\maketitle

\section{Introduction}
Recently there was a growing interest, in the Optics community,
 towards those optical fields possessing unusual
polarization features; this because some phenomena have shown to be strongly
dependent by vectorial properties of the field. 
 For instance, it was experimentally demonstrated that a radially polarized
 field can be focused to spot size of extension much smaller than that
  obtainable by using a linear polarized one \cite{Dorn03}.
   Also, azimuthally and radially polarized fields \cite{Nota1} 
 have found use in many sector of Optics, like particle beam trapping,
  optical microscopy, superrisolution \cite{Novotny01}, specially when those polarization features
   are associated to a particular intensity profile
    (the so-called donut beams are an example of that) which implies the presence of a strong
     longitudinal component of the optical electric field;
    moreover it was shown that beams possessing field lines in the shape of a spiral, namely
     spirally polarized beams, are suitable to describe fields 
  with vortices in their transverse phase profile \cite{Gori01}. In the present
   paper we wish to widen this scenario, by introducing coherent beams with a non-uniform
    polarization state with an elliptical symmetry. By doing so, we wish to show how 
    some of aforesaid properties are easily extendable to other polarization's geometry and 
    to emphasize the differences among all those polarizations.
    The paper is organized as follows:
    in section 2, after having recalled briefly some
     definitions about cylindrical coordinate system, we specialize the subject to 
     elliptical symmetry; in section 3 we derive some results in paraxial optics regime by 
     applying the theory to Bessel-Gauss beams.
        
\section{Field and polarization symmetries}
Let us suppose we have a free-space beam, which propagates along a mean direction, say $z$-axis,
on starting from a source-plane on which we know the field distribution of amplitude and phase.
In this context the term non-uniform, when addressed to the field polarization 
features, implicity refers to the transverse component
 of the field analysed through a
rectangular (Cartesian) coordinate system. So we say that a polarization
is non-uniform if, on moving along a fixed direction in the $(x,y)$ plane, we see the 
vector field changing in direction when compared with the unit vector basis ${\hat{x}}$
and ${\hat{y}}$.
In reality, it should be still possible to find a new coordinate system respect to which
the field mantains its polarization state. In other words, 
 the field should be uniformly polarized respect to this 
  new coordinate system. In a generical cylindrical orthogonal reference frame
we can write the field, across the plane $z=0$, as
\begin{equation}\label{campo completo}
{\bf E}={\bf E_T}+E_z\hat z
\end{equation}
where
\begin{equation}\label{campo curv}
{\bf E_T}(q_1,q_2,0)=E_1(q_1,q_2,0) {\hat{q}_1}+E_2(q_1,q_2,0) {\hat{q}_2}
\end{equation}
indicates the trasverse field and ${\hat{q}_1} $, ${\hat{q}_2}$ and
 $z$ indicate 
the vector basis of a generical cylindrical orthogonal coordinate system, which is related 
to a Cartesian orthogonal system by the following relationships
\begin{eqnarray}\label{curv gen}
x=g_1(q_1,q_2) \nonumber\\
y=g_2(q_1,q_2)\\
z=z\nonumber
\end{eqnarray}
where $g_1$ and $g_2$ are smooth functions of $q_1$, $q_2$ variables.
Rectangular and circular ones are special cases of equations (\ref{curv gen}).
In particular, speaking about circular symmetry, recently several authors have focused their attention
 on fields polarized in a radially or azimuthally way. In almost all these works also the field (i.e the amplitude and phase of each scalar component of the optical field, 
 and not only the polarization)
  was choosen to be circularly simmetric. While this choice seems to be the most obvious and natural, anyhow it can be interesting to study the case in which this 
  {\it coupling in symmetry} is removed, a case that does not have received attention at all.
This fact is a little odd, specially if one thinks to the way in which
a radially or azimuthally circular polarized field can be obtained in practice \cite{Tidwell90}.
For instance, one of the methods to obtain a radially polarized beam is
 to combine coherently an Hermite-Gauss beam of order $(0,1)$ and an analogous beam, but of order
 $(1,0)$, both arising
  from a laser beam, on supposing that 
 amplitude and phase of the two beams are the same. There exist other metods, 
 but all of them also suppose 
 to have a very good control of the amplitude
 of the two component that one adds togheter.
Actually this is true only approximately and in practice one can consider the effect of some 
difference in amplitude between the two emerging beams, a difference which gives rise just to 
an elliptical symmetry.
In the next section we start by analysing the form of a generical field endowed with
 features that break both the circular than rectangular symmetry and we will discuss the 
 relative properties.

\subsection{Vectorial beam with elliptical symmetry}
\begin{figure}[t1]
  \centerline{\includegraphics{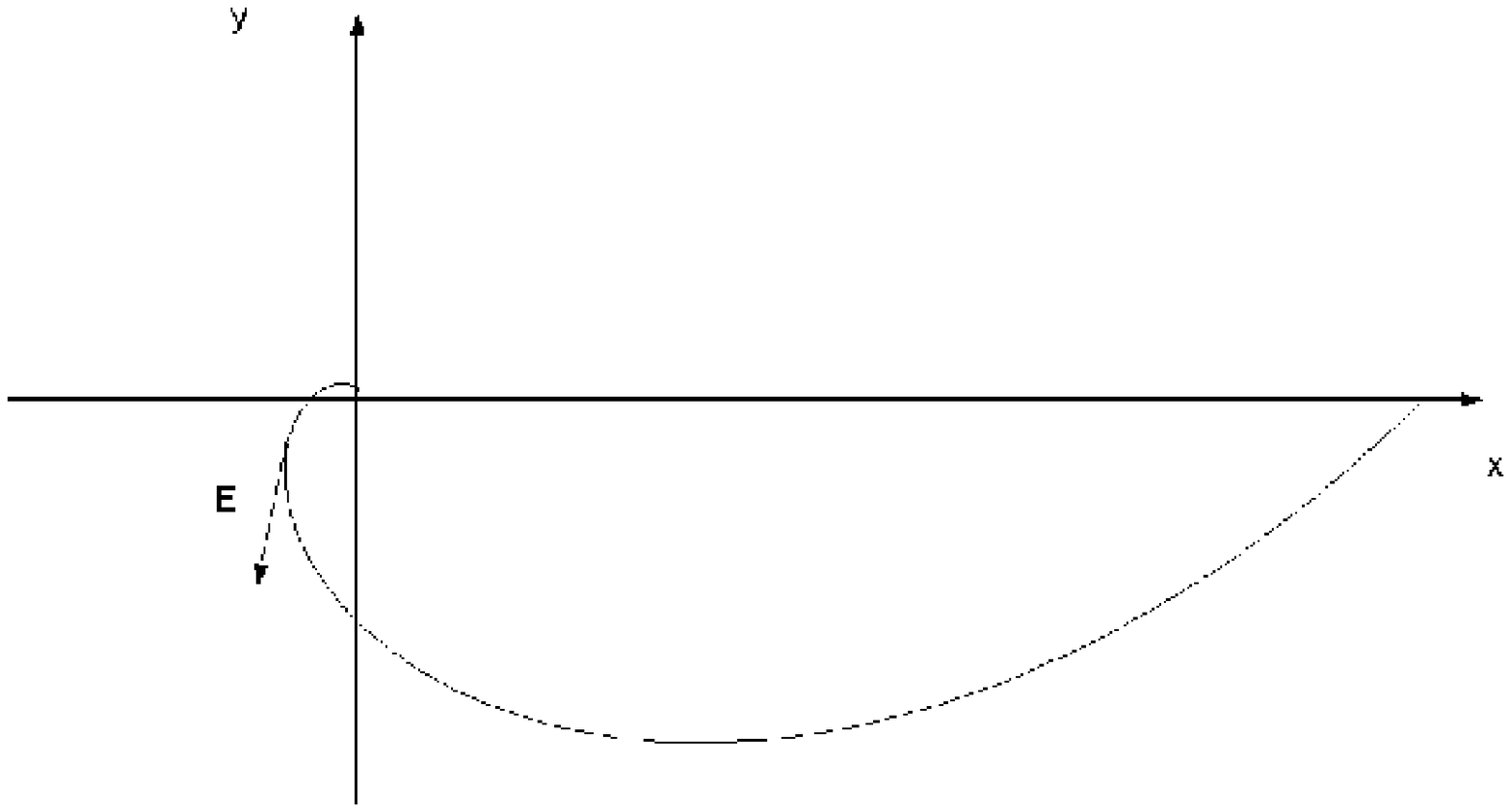}}
  \caption{elliptical logarithmic spiral. The parameters are choosen as follows: $B/A=10/3$, 
  $\alpha=0.8$.}
  \label{spiral}
\end{figure}

Let us suppose we have a field like that in equation (\ref{campo completo}) 
and ${\bf E_T}$ be
\begin{equation}\label{campo}
{\bf E_T}= A f(r,z) \cos(\theta+\alpha) {\hat x} + B f(r,z) \sin(\theta+\alpha) {\hat y}
\end{equation}
where $A$ and $B$ indicate two arbitrary real constants, $r$ and $\theta$ are, respectively,
 the radial and angular coordinates of a circular-symmetric coordinate system, 
$f(r,z)$ is a radial-circular function across a $z$ plane and $\alpha$ is a constant angle.
We suppose here that $A\neq B$. 
It is straightforward to show that the field in equation (\ref{campo}) can be rewritten as follows
\begin{equation}\label{campo ell} 
{\bf E_T}= C f(r,z) [\sinh\xi \cos(\theta+\alpha){ \hat x}
+ \cosh\xi\sin(\theta+\alpha) {\hat y}]
\end{equation} by choosing the terms $C$ and $\xi$ as
\begin{equation}\label{Cambio 1}
C^2=B^2-A^2
\end{equation}
\begin{equation}\label{Cambio 2}
\cosh\xi=\frac{B}{C}
\end{equation}
The first observation we do is trivial. If we fix the radial coordinate and we let to 
vary only $\theta$-variable, we will see the arrow describing the electric field in equation 
(\ref{campo})
to describe an ellipse with semiaxis $A f(r,z)$ and $B f(r,z)$. 
More interesting is to find the shape of the field lines for the aforesaid field. To achieve this
 goal we prefer to refer to a circular coordinate system. To obtain the field lines one needs to 
 solve the equation
\begin{equation}\label{field lines1}
{\frac{dy}{dx}}={\frac{E_y}{E_x}}={\frac{B \sin(\theta+\alpha)}{A \cos(\theta+\alpha)}}
\end{equation}
which becomes, in polar coodinates,
\begin{equation}\label{field lines2}
{\frac{\sin\theta dr+r\cos\theta d\theta}{\cos\theta dr-r\sin\theta d\theta}}=
a{\frac{\sin\theta\cos\alpha+\cos\theta\sin\alpha}{\cos\theta\cos\alpha-\sin\theta\sin\alpha}}
\end{equation}
where we have defined a new constant $a=B/A$.
Equation (\ref{field lines2}) can be transformed into the following relation
\begin{equation}\label{field lines3}
{\frac{dr}{r}}=d\theta{\frac{[-(\cos^2\theta+a\sin^2\theta)\cos\alpha+\sin2\theta\sin\alpha(1-a)/2]}
{[\sin2\theta\cos\alpha(1-a)/2-\sin\alpha(\sin^2\theta+a\cos^2\theta)]}}
\end{equation}
by recalling the relationship between Cartesian and circular coordinates.
We note that when $a=1$ (i.e in the circular-symmetric case) this equation reduces to
\begin{equation}\label{field lines circ}
{\frac{dr}{r}}=d\theta{\frac{\cos\alpha}{\sin\alpha}}
\end{equation}
and the solution is a line in the shape of a logarithmic (circular) spiral \cite{Gori01}.
Solving equation (\ref{field lines2}) is a little more difficult because the presence 
of the coefficient $a$ breaks the polar symmetry. 
We can start by studying two particular cases, i.e, when $\alpha=0$ and $\alpha=\pi/2$.
In the first case ($\alpha=0$) equation (\ref{field lines2}) becomes
\begin{equation}\label{field lines 0}
{\frac{dr}{r}}=d\theta{\frac{\cos^2\theta+a\sin^2\theta}{(a-1)\sin\theta \cos\theta}}
\end{equation}
This equation can be solved, by rewriting it in the following form
\begin{equation}\label{field lines 01}
{\frac{dr}{r}}=d\theta{\frac{\cos\theta}{(a-1)\sin\theta}}+{\frac{a\sin\theta}{(a-1)\cos\theta}}
\end{equation}
and it gives
\begin{equation}\label{lines 0}
r(\theta)=R[{\frac{\sin\theta}{(\cos\theta)^a}}]^{1/(a-1)}
\end{equation}
where $R$ is an integration constant. Equation (\ref{lines 0}) describes open lines, which tend 
to lines through the origin when $a \rightarrow 1$, and that
represent the analogous of radial-circular polarization for the geometry in subject, namely a 
radial-elliptical polarization. In Cartesian coordinates it reduces to the 
following relation between $x$ and $y$ coordinates
\begin{equation}\label{lines 0cartesian}
y=K x^a
\end{equation}
where K is a constant value.
In the other case ($\alpha=\pi/2$) equation (\ref{field lines3}) becomes
 \begin{equation}\label{field lines pi/2}
{\frac{dr}{r}}=d\theta(a-1){\frac{\sin\theta\cos\theta}{(1+(a-1)\cos^2\theta}}
\end{equation} 
and it gives
 \begin{equation}\label{lines pi/2}
r(\theta)=R{\frac{1}{(1+(a-1)\cos^2\theta)^{1/2}}}
\end{equation}
This is the polar form of ellipses, with (on considering $a>1$) the major axis parallel 
to y-axis and it represents the azimuthal polarization for elliptical geometry namely 
an azimuthal-elliptical polarization. When $a=1$ the ellipses become circles (described by 
$r=R$ relation), i.e. we find again 
the circularly-symmetric case.
For the intermediate case, that is for $\alpha\neq0,\pi/2$, the integration of 
the respective differential equation is less easy. In Figure~(\ref{spiral}) we report the solution after having choosen $\alpha=0.8$ and the quotient $a=B/A=10/3$. The field line that arises is a 
 logarithmical (but elliptical) spiral. Spirally
 beams were introduced by Gori \cite{Gori01} for circular-symmetric case and here
 we found a natural extension to elliptically-symmetric case.   
In the following we 
 analyse some properties of this kind of beam.
We have already seen which is the role of $\alpha$ angle: when it tends to zero 
the polarization tends to become radial-elliptical, 
the field being tangent to curves like in equation (\ref{lines 0cartesian}) whereas when it
tends to $\infty$ the polarization tends to be azimuthal-elliptical, the field being
tangent to ellipses described by equation (\ref{lines pi/2}).
Also we note that the beam described by equation (\ref{campo ell}) is anisotrope,
 i.e its tranverse intensity pattern
depends by $\theta$ angle. Indeed, for the intensity of transverse component holds
\begin{equation}\label{I}
I_T(r,\theta,z)\propto{|\bf E_T|}^2=C^2|f(r,z)|^2[\cosh^2\xi-\cos^2(\theta+\alpha)]
\end{equation}
in which the angular dependence is evident. It is simple to see 
that the maximum intensity directions are
\begin{equation}\label{max dir}
\theta_1=\pi/2-\alpha\pm m\pi\nonumber\\
\end{equation}
($m=0,1,2..$) in which equation (\ref{I}) becomes 
\begin{equation}\label{I max}
I_T(r,\theta,z)\propto{|\bf E_T|}^2=C^2|f(r,z)|^2 \cosh^2\xi
\end{equation} 
so we can find the value of angle $\alpha$ and, 
as a consequence, also the polarization kind, by intensity measurement,
 contrary to the circular-symmetric case in which 
 the intensity is dependent only from the radial coordinate $r$ and does not give any
  information about polarization. In other words, in the circular-symmetric case
  all transverse intensity profiles associated to a particular
   polarization state (radial, azimuthal, spiral ones, etc.) are
  undistinguishable one to another. In the elliptical-symmetric case, one can obtain
   all the parameters, i.e. the ratio $B/A$ and the angle $\alpha$, simply by the 
   intensity transverse map measured across a generical tranverse plane. Indeed, with reference to
   equation (\ref{I}), by measuring the maximum intensity direction one evalues
    the angle $\alpha$ (equation (\ref{max dir}), whereas by measuring the angular-visibility ratio 
   \begin{equation}\label{visib}
   \nu_r=\frac{I_{Max}-I_{Min}}{I_{Max}}=\frac{1}{\cosh^2\xi}
   \end{equation}
    one evalues the values of $A$ and $B$ 
    (the indice $r$ indicates that the radial coordinate is fixed).
   Also we note that the equation (\ref{I}) never goes to zero,
   this because the term $\cosh^2\xi-\cos^2(\theta+\alpha)$ 
is different to zero $\forall \xi\neq0$ (the case $\xi=0$ represents linear polarization and 
here is not considered). It follows that it ever holds $\nu<1$. 
It remains to see how one can obtain such a field. This is the purpose of the next
section.
\section{Elliptically Symmetric Polarized Bessel-Gauss Beams}
\begin{figure}[t2]
  \centerline{\includegraphics{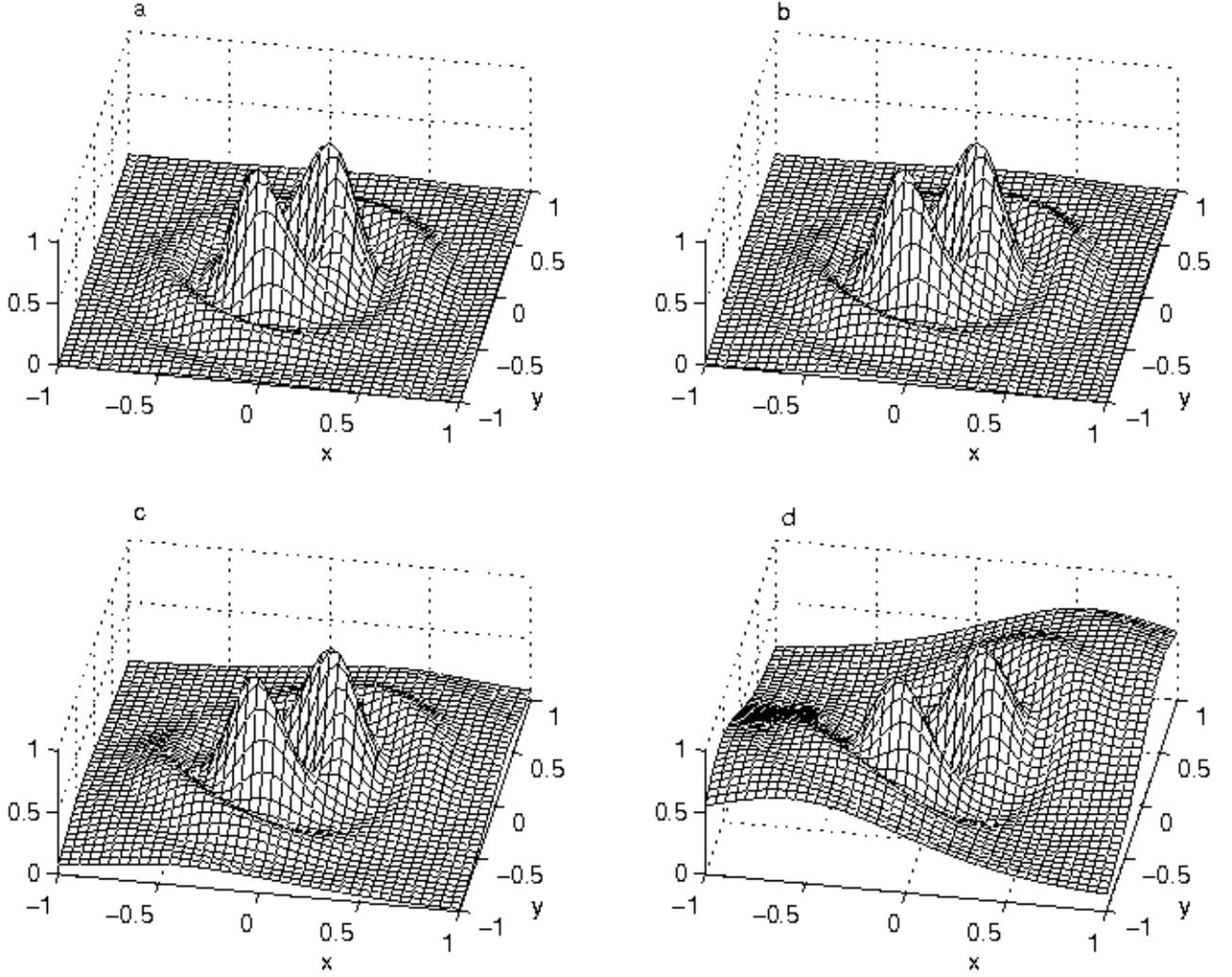}}
  \caption{Tranverse intensity of 
  a  ESP Bessel-Gauss beam with non-uniform elliptical polarization at different $(x,y)$ plane.
  a) $z=0$; b)$z=0.1L$;3)$z=0.2L$;4)$z=0.3L$, where $L$ is Rayleigh length. The intensity
  is  normalized at one to each plane.
  With reference to equation(\ref{ESPB}),
   we have $\beta=8/w$, $\alpha=0.8$, $B/A=2$. The axises are normalized to $w$ value.}
  \label{Intensita}
\end{figure}

\begin{figure}[t3]
  \centerline{\includegraphics{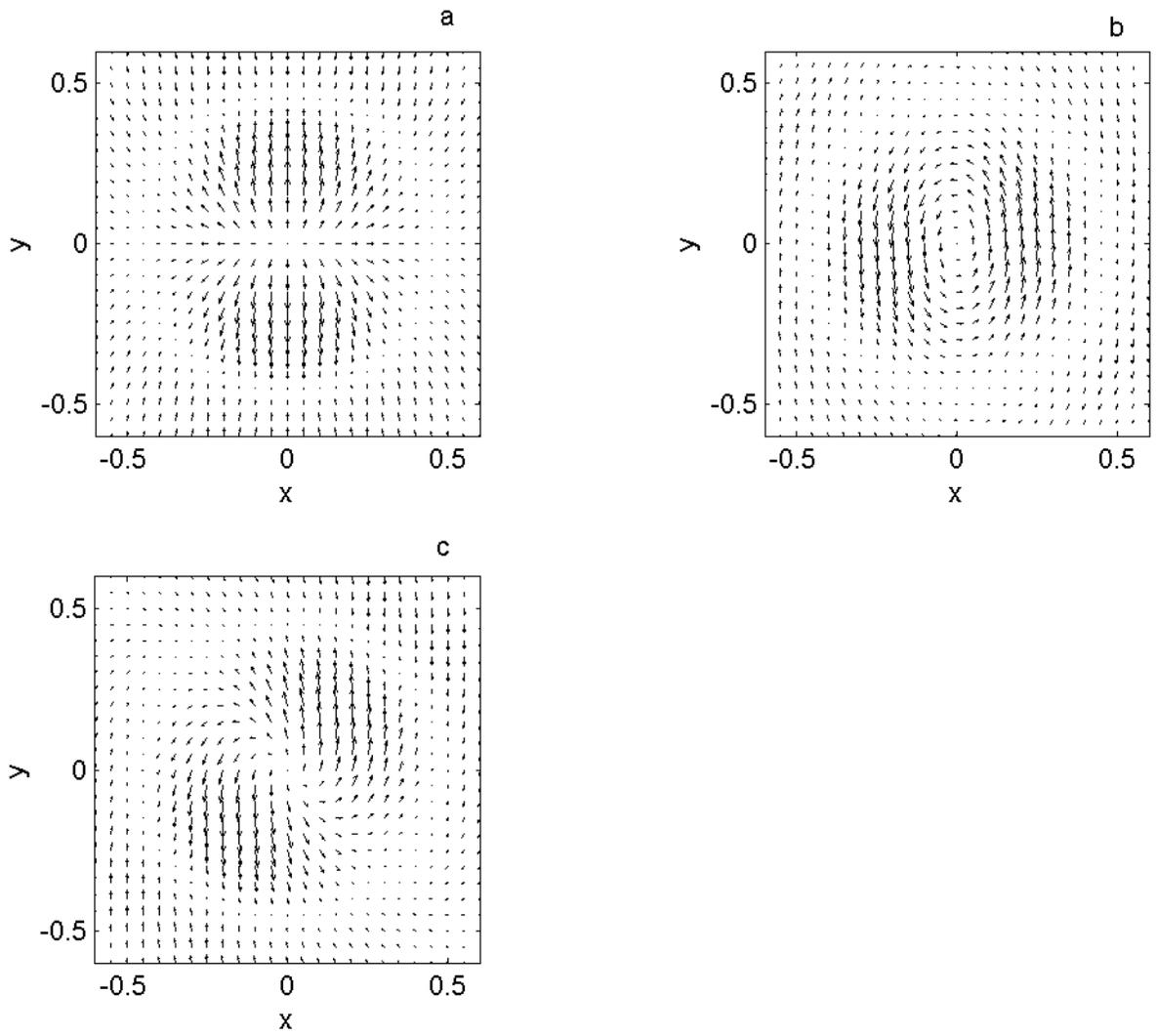}}
  \caption{Polarization pattern of a ESP Bessel-Gauss beam across source plane ($z=0$). The parameters
  are choosen as follows:  $B/A=10/3$,$\beta=8/w$ and a)$\alpha=0$, b)$\alpha=\pi/2$, c)$\alpha=0.8$. 
  The axises are normalized to $w$ value.}
  \label{Pol patt}
\end{figure}

\begin{figure}[t5]
  \centerline{\includegraphics{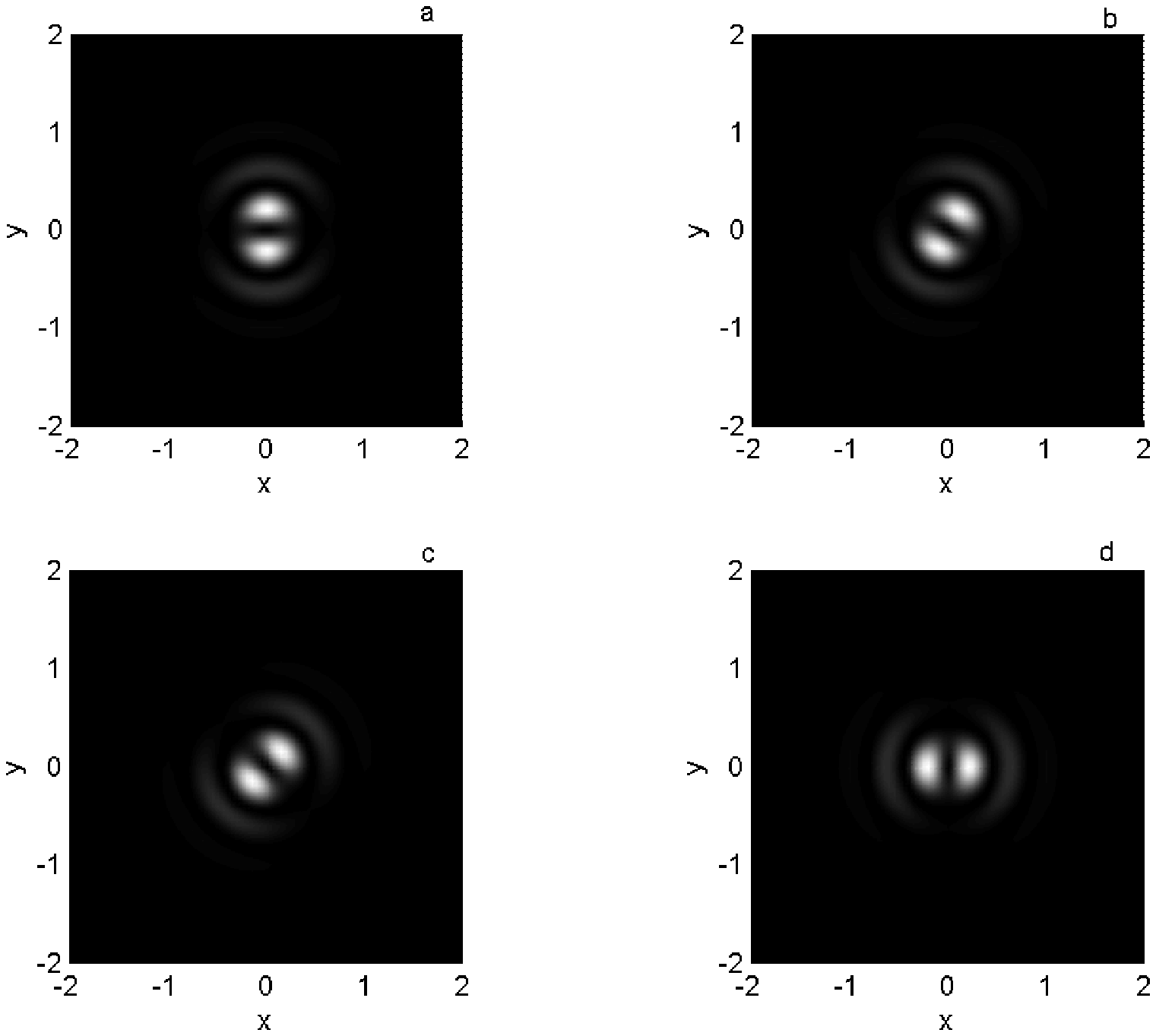}}
  \caption{Intensity map of a ESP Bessel-Gauss beam
   across source plane ($z=0$) for different
   values of polarization angle $\alpha$: a) $\alpha=0$, b) $\alpha=\pi/6$ 
    c) $\alpha=\pi/4$ d)$\alpha=\pi/2$. The parameters
  are choosen as follows:  $B/A=10/3$,$\beta=8/w$. The axises are normalized to $w$ value.}
  \label{vari alfa}
\end{figure}

It is well known that Bessel-Gauss beams ($BG$ beams for short) are solution of scalar
paraxial wave equation and were introduced for the first time from 
{Gori \it et al.} \cite{Gori87}.
In particular Gori introduced the linearly polarized $BG$ beams of zero-order, here
denoted for convenience as $BG_0$. Subsequently many authors have been interested
both to the generalization to vectorial \cite{Greene98} and higher-order
 $BG$ beams, both to beam with
 unusual polarization features, as azimuthal and radial ones \cite{Jordan94}. Here we recall,
for reader convenience, the form of a $BG_0$
\begin{equation}
BG_0= \frac{1}{(1+iz/L)}J_0[\frac{\beta r}{(1+iz/L)}]\exp[\frac{-r^2}{w^2 (1+iz/L)}]
\exp[\frac{-i\beta^2 z}{2k(1+iz/L)}]
\end{equation}
where $J_0$ is the zero-order Bessel function of the first kind, $L$ is the Rayleigh length, $w$
is the minimum spot size of the Gaussian beam, and $\beta$ is the intensity of the 
tranverse wave vector;
to extend the vectorial treatise to encompasse also a non-uniform elliptical
polarization we start from a vectorial paraxial wave equation.
On starting from Maxwell's equations in vacuum, in absence of sources, we arrive
to the following equation for the complete vector field $\bf E$
\begin{eqnarray}
\nabla\times\nabla\times{\bf E}=k^2{\bf E}\nonumber\\
\nabla\cdot{\bf E}=0
\end{eqnarray}
On introducing a "slowly varying part" of ${\bf E}$, say ${\bf F}$, with
\begin{equation}
{\bf E}={\bf F}\exp(ikz)
\end{equation}
and by making use of the paraxial approximation, 
we obtain that ${\bf F}$ is governed from the following equations \cite{Borghi04}
\begin{equation}\label{parax vett}
\nabla^2_T{\bf F_T}+2ik\partial_z{\bf F_T}=0
\end{equation}
\begin{equation}\label{parax vett long}
\nabla_T\cdot{\bf F_T}+ikF_z=0
\end{equation}
in which we have decomposed the field ${\bf F}$ in longitudinal and transverse component, i.e.
${\bf F}={\bf F_T}+F_z \hat z$.
Now, if we decompose also the field $ {\bf F_T}$ into rectangular component, i.e.
we let ${\bf F_T}=F_x \hat x+F_y\hat y$ we obtain two scalar paraxial wave equations, each for single
component,
\begin{equation}\label{Parax 1}
\nabla^2_T{ F_x}+2ik\partial_z{F_x}=0
\end{equation}
\begin{equation}\label{Parax 2}
\nabla^2_T{ F_y}+2ik\partial_z{ F_y}=0
\end{equation}
equation (\ref{Parax 1})-(\ref{Parax 2}) can be rewritten as $LF_x=0$ and $LF_y=0$, after 
introduced a linear differential operator $L=\partial^2_x+\partial^2_y+2ik\partial_z$.
Now, as the $L$ operator commutes with the infinitesimal displacement operators \cite{Wunsche89}
$\partial_x $, $\partial_y$ and $\partial_z$, it is possible to obtain new solutions
from known solutions by mean of arbitrary differentiation. This observation will be crucial
to obtain elliptically symmetric polarized $BG$ beams. Indeed, the first thing
to do is differentiating, respect to $x$ and $y$ variables,
 a $BG_0$ beam;  in this way we obtain two new possible solutions for a scalar paraxial
wave equation: $\partial_x BG_0$ and $\partial_y BG_0$.
If we now linearly combine them properly we find the ESPB. To  
this purpose it is sufficient to let
\begin{eqnarray}\label{combinazioni lin 1}
F_x=A[\partial_x(BG_0) \cos\alpha-\partial_y(BG_0)\sin\alpha]\\
F_y=B[\partial_y(BG_0) \cos\alpha+\partial_x(BG_0)\sin\alpha]
\end{eqnarray}
in which $A$ and $B$ are two arbitrary constant, with $A\neq B$, and $\alpha$, as usual, 
is a constant angle. On utilizing the position in equations
(\ref{Cambio 1})-(\ref{Cambio 2}) we finally obtain
\begin{eqnarray}\label{ESPB}
{\bf F_T}({\bf r},z) & = &  \frac{C}{(1+iz/L)^2}\exp[-\frac{r^2}{w^2 (1+iz/L)}] \times \nonumber \\
& \times & \exp[-\frac{i\beta^2 z}{2k(1+iz/L)}][\beta J_1(\frac{\beta r}{1+iz/L})+\nonumber\\
& + & \frac{2r}{w^2}J_0(\frac{\beta r}{1+iz/L})][\sinh\xi \cos(\theta+\alpha) \hat x+\cosh\xi \sin(\theta+\alpha) \hat y]
\end{eqnarray}
As the paraxial propagation has to be satisfied, from equation (\ref{ESPB}) follows that the function $f(r,z)$ on the source plane ($z=0$)
 can't be choosen in a arbitrary way. In particular, by comparing equation (\ref{campo}) and equation (\ref{ESPB}),
  we have
\begin{equation}\label{campo0}
f(r)=f(r,z)|_{z=0}=\exp[-\frac{r^2}{w^2}]
[\beta J_1(\beta r)+\frac{2r}{w^2}J_0(\beta r)]
\end{equation}
In Figures~(\ref{Intensita}) we report the intensity
 pattern of a ESP Bessel-Gauss beam, evalued at different transverse planes.
  We see that the beam mantains, during its paraxial propagation, a zero 
 in correspondence of $z$ axis (a properties that is shared 
 with the classical donut beams). In Figure~(\ref{Pol patt}) we report the polarization pattern,
  in correspondence to the $z=0$ plane, 
 in which is shown the behaviour of tranverse field for $\alpha=0.8$, and a ratio $B/A=10/3$.
 On changing the sign of the linear combination in equation (\ref{combinazioni lin 1})
   as well as the ratio $B/A$ one can obtain other interesting polarization pattern.
    In Figure~(\ref{vari alfa}) we plot intensity across the tranverse plane $z=0$, for 
    different values of angle $\alpha$, to show the aforesaid angular dependence.
Before closing the present section, we wish to say something about the longitudinal component
of the electric field. To do that we will refer to the generical form 
 of EPSB as that in equation (\ref{campo ell}) and we shall use 
 equation (\ref{parax vett long}) to obtain the longitudinal component in paraxial regime:
\begin{equation}\label{Fz}
F_z= \frac{i}{k}\nabla_T\cdot{\bf F_T}
\end{equation}
 After straightforward calculations we find
\begin{eqnarray}\label{long}
F_z & = & (\partial_r{f(r,z)}+\frac{f(r,z)}{r})\sinh\xi \cos\alpha +\exp(-\xi)[(\partial_r{f(r,z)}\sin^2\theta+\\\nonumber
& + & \frac{f(r,z)}{r} \cos^2\theta)\cos\alpha+\frac{1}{2}(\partial_r{f(r,z)}-\frac{f(r,z)}{r})\sin\alpha \sin2\theta]
\end{eqnarray}
which represents the full form of the longitutinal component in paraxial approximation.
Here we see that, contrary to the circular-symmetric case for which an azimuthally 
polarized field does not give rise to a component along the z-axis, 
when we have an azimuthal-elliptical
polarization (for which $\alpha=\pi/2$), $F_z$ is not null and holds
\begin{equation}\label{long 2}
F_z= \exp(-\xi)[\frac{1}{2}(\partial_r{f(r,z)}-\frac{f(r,z)}{r})\sin2\theta]
\end{equation}
and it goes to zero as $\exp(-\xi)$. This is obvious if we recall that when $\xi \rightarrow \infty$
the geometry polarization tends to become circular because for 
high value of $\xi$, $\sinh\xi \approx \cosh\xi \approx \exp(\xi)$ and the ellipses in 
 equation (\ref{lines pi/2}) tend to become circles.
The fact that the longitudinal component changes by changing coordinate system is well-known 
 and it can explained by mean a geometrical point of view. Indeed, equation (\ref{Fz}) states that 
${F_z}$ is a divergence of a vectorial field. The divergence $\nabla_T\cdot{\bf F_T}$
 counts the algebrical sum of sinks (i.e. where a field line of ${\bf F_T}$ terminates) 
 and sources (i.e. where a field line of ${\bf F_T}$ originates) 
 into a surface element, across a $(x,y)$ plane. 
Since the divergence is a pseudoscalar, namely we are dealing with a {\it density},
it is not invariant under a generical space deformation.
If, for example, the space is squeezed in a way in which the areas are reduced, 
the density (i.e. $F_z$) increases.
In practice, a distortion of space can be thought as a 
transformation of coordinates and the fact 
that the divergence operation is dependent by the coordinate 
system in which we work means that we are dealing with metrical, and not topological, features
of the field \cite{Weinreich}.

\section{Conclusion}
We have introduced unconventionally polarized optical beams 
endowed with elliptically-symmetric features in their 
tranverse polarization pattern.
 A beam of this kind mantains its polarization state under free-space paraxial propagation. 
 In particular we have derived a natural extension to 
elliptically-symmetric case
 of spirally-polarized beams introduced by Gori \cite{Gori01} for circular symmetric
 case. 
  Some properties of this kind of field have been analysed. In particular,
   we have pointed out that, by using an 
  elliptically-symmetric polarization, the tranverse intensity profile 
  gives directly information about
   the polarization state, the angular behaviour of the intensity depending from it.
   This implies that it is possible to measure the field polarization state only by an 
   acquisition of the intensity pattern across a generical tranverse plane. This is in 
   contrast with the analogous situation in circular-symmetric case, in which the intensity
    pattern is indipendent from the particular polarization state. Indeed, a field polarized
     in a radially way produces a tranverse intensity profile identical to that produced 
     by the azimuthally or spirally polarized ones,
      or any other combination of them which mantains the circular symmetry.
Due to its important role in optics microscopy, we have reported also the form of the longitudinal 
component for the electric field, in paraxial approximation.

\section{Acknowledgement}
The author wishes to thank Massimo Santarsiero for useful discussions and suggestions about the subject 
of the present work.



\begin{thebibliography}{99}
\bibitem{Nota1}
As it will be more clear in the following, 
it would be more correct to refer to such polarization
feature as azimuthal-circular and radial-circular one in order that one can 
put into evidence the coordinate system to which they refer.
\end{thebibliography}
\end{document}